\documentclass[twocolumn,floatfix,
superscriptaddress,citeautoscript,
amsmath,amssymb,amsfonts,aps,prb]{revtex4-1}
\usepackage{graphicx}
\usepackage{xcolor}
\usepackage{soul}
\usepackage{newtxtext}
\usepackage{booktabs}
\usepackage[cmintegrals]{newtxmath} 
\usepackage{graphicx}
\usepackage{dcolumn}
\usepackage{bm}
\usepackage{hyperref}
\usepackage{amsmath}
\usepackage{enumitem}
\usepackage{cleveref}
\hypersetup{colorlinks,
	linkcolor={blue!75!black!80!yellow},
	citecolor={blue!75!black!80!yellow},
	urlcolor={blue!75!black!80!yellow}
}

\makeatletter \renewcommand\@make@capt@title[2]{%
\@ifx@empty\float@link{\@firstofone}{\expandafter\href\expandafter{\float@link}}%
\sffamily{\textbf{#1}}\@caption@fignum@sep#2 }

\usepackage[normalem]{ulem}
\makeatletter
\renewcommand\@make@capt@title[2]{%
    \@ifx@empty\float@link{\@firstofone}{\expandafter\href\expandafter{\float@link}}%
    \sffamily{\textbf{#1}}\@caption@fignum@sep#2
}%

\begin{document}

\title{Probing carrier interactions using electron hydrodynamics}

\author{Georgios Varnavides}
\email{gvarnavi@mit.edu}
\thanks{These authors contributed equally.}
\affiliation{Department of Materials Science and Engineering, Massachusetts Institute of Technology, Cambridge, MA, USA}
\affiliation{Research Laboratory of Electronics, Massachusetts Institute of Technology, Cambridge, MA, USA}
\affiliation{John A. Paulson School of Engineering and Applied Sciences, Harvard University, Cambridge, MA,USA}

\author{Adam S. Jermyn}
\email{adamjermyn@gmail.com}
\thanks{These authors contributed equally.}
\affiliation{Center for Computational Astrophysics, Flatiron Institute, New York, NY 10010, USA}

\author{Polina Anikeeva}
\affiliation{Department of Materials Science and Engineering, Massachusetts Institute of Technology, Cambridge, MA, USA}
\affiliation{Research Laboratory of Electronics, Massachusetts Institute of Technology, Cambridge, MA, USA}

\author{Prineha Narang}
\email{prineha@seas.harvard.edu}
\affiliation{John A. Paulson School of Engineering and Applied Sciences, Harvard University, Cambridge, MA,USA}

\date{\today} 

\begin{abstract}

Electron hydrodynamics arises when momentum-relaxing scattering processes are slow compared to momentum-conserving ones.
While the microscopic details necessary to satisfy this condition are material-specific, experimentally accessible current densities share remarkable similarities.
We study the dependence of electron hydrodynamic flows on the rates of momentum-relaxing and momentum-conserving scattering processes in a microscopics-agnostic way.
We develop a framework for generating random collision operators which respect crystal symmetries and conservation laws and which have a tunable ratio between the momentum-conserving and momentum-relaxing lifetimes.
Using various random instances of these collision operators, we calculate macroscopic electron viscosity tensors and solve the Boltzmann transport equation (BTE) in a channel geometry over a grid of momentum-conserving and momentum-relaxing lifetimes, and for different crystal symmetry groups.
We find that different random collision operators using the same lifetimes produce very similar current density profiles, meaning that the current density is primarily a probe of the overall rates of momentum conservation and relaxation.
By contrast, the viscosity tensor varies substantially at fixed lifetimes, meaning that properties like channel resistance provide detailed probes of the underlying scattering processes.
This suggests that, while details of the scattering process are imprinted in the electronic viscosity tensor, for many applications theoretical calculations of hydrodynamic electron flows can use experimentally-available lifetimes within a spatially-resolved BTE framework rather than requiring the costly computation of \textit{ab initio} collision operators.

\end{abstract}

\maketitle

\section{Introduction}\label{sec:intro}

Spatially-resolved experiments have revealed that electrons in condensed matter can flow collectively akin to classical fluids~\cite{KrishnaKumar2017,sulpizio_visualizing_2019,Marguerite2019,Ku2020,jenkins_imaging_2020}, confirming theoretical predictions over fifty years old~\cite{Gurzhi_1968}.
Such electron ``hydrodynamics'' is observed in the limit where microscopic momentum-relaxing interactions, for example, scattering of electrons against impurities and/or lattice vibrations, are slow compared to momentum-conserving electron-electron interactions, such as the repulsive Coulomb interaction.

Since electron screening effects are expected to minimize direct (Coulomb) electron-electron interactions in bulk conductors, the seemingly-serendipitous observations of electron hydrodynamics in bulk (semi-)metals~\cite{moll2016evidence,Gooth2018thermal,Osterhoudt2021evidence,vool2021imaging,aharonsteinberg2022direct} have garnered significant attention, and recent work attributes these to indirect lattice-mediated electron interactions~\cite{vool2021imaging,varnavides2021finitesize,wang2021generalized}.
Despite the differences in the microscopic origins of momentum-conserving interactions, the hallmark features of channel or ``Poiseuille'' flow, that is, enhanced current density in the center of the channel and reduced current density at the edges~\cite{sulpizio_visualizing_2019,Ku2020,jenkins_imaging_2020,vool2021imaging}, appear similar.

At the same time, the theoretical methods used to investigate these electron hydrodynamic flows often rely on simplifying models, such as the electronic Stokes equations~\cite{succi2011,Levitov2016,superballistic2017,varnavides2020electron,scaffidi2017,schmalian2019} and the dual relaxation time approximation of the Boltzmann transport equation~\cite{Heremans2004,Cepellotti2015,stokesParadox2016,scaffidi2017,levitov2019,schmalian2019,Callaway1959,guyer1966,deJong1995}.
The former is only strictly valid in the unphysical limit of zero momentum relaxation, while the latter effectively convolves all microscopic details to arrive at scalar lifetimes which greatly simplify the kinetic theory used.

More recently, numerical methods to solve the linearized Boltzmann transport equation iteratively with spatial-resolution have been developed and applied to study plasmonic hot carriers~\cite{jermyn-prmat}, and phonon transport~\cite{varnavides-prb,romano2020phonon}.
While these methods allow one to go beyond the dual relaxation time approximation, it is computationally expensive to calculate the linearized collision operator for electron interactions from first principles.

Taken together, these observations pose questions that are key to our understanding of electron hydrodynamics:
\begin{enumerate}[label=\alph*)]
    \item How sensitive are \textit{macroscopic} observables of electron hydrodynamic flows to \textit{microscopic} interaction details?
    \item Can we identify the linearized collision \textit{operator} using experimentally-accessible \textit{scalar} interaction lifetimes?
\end{enumerate}

In this \emph{Article}, we investigate these questions statistically.
In~\cref{sec:form} we propose a procedure to construct \textit{physically-plausible random} linearized collision operators using the crystal symmetry of the system in question and conservation laws of properties such as eigen-energies, group velocities, and scalar interaction lifetimes.
The latter can be obtained using temperature-dependent first principle calculations~\cite{coulter2018,garcia2021,varnavides2021finitesize,wang2021generalized}, or extracted from transport measurements~\cite{Gooth2018thermal,vool2021imaging}.
In~\cref{sec:results} we use the inherent \textit{randomness} in the proposed procedure as a proxy for the differences introduced by the mechanism- and material-specific microscopic interaction details, and evaluate their variability on macroscopic observables, such as current density measurements in two-dimensional channel flow.

Two key findings emerge from our analysis:
First, different random instances of the collision operator with the same interaction lifetimes produce very similar current density profiles.
That is, the details of the scattering processes at work in a material \textit{do not} set the current density; rather it is controlled by the overall rates of momentum conservation and relaxation. As a result experimental measurements of the current density profile probe these overall rates, not the details of the scattering processes at work.
Secondly, the electron fluid's viscosity tensor is sensitive to microscopic details of the collision operator.  Its components vary by more than $\sim$50\% between different collision operators. Thus the action of viscosity on the channel flow resistance provide an experimental probe of the underlying scattering processes.
These observations are consistent across different crystal symmetry groups, and provide quantitative estimates of the error introduced by the dual relaxation time approximation.

Finally, we note that for most applications, the error introduced by approximating the linearized collision operator with experimentally-available lifetimes is within the experimental uncertainty of current profile measurements~\cite{sulpizio_visualizing_2019,Ku2020,jenkins_imaging_2020,vool2021imaging,aharonsteinberg2022direct}, and provides a promising efficient alternative to \textit{ab-initio} calculations.
\section{Formalism}\label{sec:form}
\subsection{Boltzmann Transport Equation}\label{sec:bte}

We consider the general transport problem given by the semi-classical Boltzmann transport equation (BTE):
\begin{align}
	\partial_t n_i(\boldsymbol{r}) + \boldsymbol{v}_i\cdot\nabla_{\boldsymbol{r}} n_i(\boldsymbol{r}) = -\boldsymbol{F}\cdot\nabla_{\boldsymbol{k}} n_i(\boldsymbol{r}) - \Gamma_i[\{n_j(\boldsymbol{r})\}],
	\label{eq:bte1}
\end{align}
where $n_i(\boldsymbol{r})$ is the distribution function of carriers in combined state index $i$ (encompassing the wavevector $\boldsymbol{k}$ and band $\nu$), $\boldsymbol{v}_i$ is the group velocity of state $i$, $\boldsymbol{F}$ is an external driving force, and $\Gamma_i[\{n_j(\boldsymbol{r})\}]$ is the collision operator, which specifies the rate at which carriers scatter into and out of state $i$ as a function of the full carrier distribution $n_j(\boldsymbol{r})$ at position $\boldsymbol{r}$.
A key observation underpinning our approach is that the collision operator is local in space, while the advection operator $\boldsymbol{v}_i\cdot\nabla_{\boldsymbol{r}}$ is local in state-space.

Our framework rests on several assumptions.
First, we take the carrier distribution to be in steady state, eliminating the first term in equation~\eqref{eq:bte1}.
Next, we linearize the collision operator about an equilibrium distribution $n^0_j$, with
\begin{align}
	\delta n_j \equiv n_j - n^0_j
\end{align}
such that
\begin{align}
	\Gamma_i[\{n_j(\boldsymbol{r})\}] \approx \frac{\partial \Gamma_i}{\partial n_j}\delta n_j,
\end{align}
where summation is implied over repeated indices.
For fermions we take $n_j^0$ as the Fermi-Dirac distribution, while for bosons we take it as the Bose-Einstein distribution.
Finally, we assume that the temperature and material properties are spatially uniform, such that $n^0_j$ is not a function of position and
\begin{align}
    -\boldsymbol{F}\cdot\nabla_{\boldsymbol{i}} n_i(\boldsymbol{r}) \approx -\boldsymbol{v}_i \cdot \frac{\partial n_i^0}{\partial \epsilon_i} \boldsymbol{F} = S_i,
\end{align}
where $\epsilon_i$ is the eigen-energy of state $i$, and we have identified the linearized forcing terms, which depend only on the equilibirum distribution, with a `source' term $S_i$ of carriers in state $i$.
With these assumptions, the BTE simplifies to
\begin{align}
	\boldsymbol{v}_i\cdot\nabla_{\boldsymbol{r}} \delta n_i(\boldsymbol{r}) = S_i - \frac{\partial \Gamma_i}{\partial n_j}\delta n_j(\boldsymbol{r}).
	\label{eq:bte2}
\end{align}

\Cref{eq:bte2} can be re-written in the form:
\begin{align}
    \left(\frac{\partial \Gamma_i}{\partial n_j} + \delta_{ij} \boldsymbol{v}_i \cdot \nabla_{\boldsymbol{r}} \right)\delta n_j(\boldsymbol{r}) = S_i,
\end{align}
which highlights that the solution $\delta n_j(\boldsymbol{r})$ may be obtained by inverting the operator in parentheses.
However, the operator to be inverted exists over the large joint space of spatial and state dimensions, so we proceed iteratively~\cite{jermyn-prmat,varnavides-prb}.
First, we separate the collision operator into diagonal terms, representing decay with lifetime $\tau_i$, and off-diagonal `mixing' terms:
\begin{align}
    \frac{\partial \Gamma_i}{\partial n_j} = \tau_{i}^{-1}\delta_{ij} - M_{ij} \label{eq:bte-decomp}.
\end{align}
Using this decomposition the BTE may be written as:
\begin{align}
	(1 + \tau_i\boldsymbol{v}_i\cdot\nabla_{\boldsymbol{r}}) \delta n_i(\boldsymbol{r}) = \tau_i S_i + \tau_i M_{ij}\delta n_j,
	\label{eq:bte3}
\end{align}
where the left-hand and right-hand sides contain only terms local in state- and position-space respectively.

We next express $\delta n_j$ as a power series in the matrix 
\begin{align}
    G_{ij} &\equiv \tau_i M_{ij}, \notag
\end{align}
where no summation is implied, such that
\begin{align}
	\delta n_j = \delta n_j^0 + \delta n_j^1 + ...
\end{align}
where
\begin{align}
	(1 + \tau_i\boldsymbol{v}_i\cdot\nabla_{\boldsymbol{r}}) \delta n_i^0 & = \tau_i S_i \\
	(1 + \tau_i\boldsymbol{v}_i\cdot\nabla_{\boldsymbol{r}}) \delta n_i^k &= G_{ij}\delta n_j^{k-1}.
\end{align}
This approach converges so long as the spectral radius of $G_{ij}$ is less than unity.
Otherwise, more sophisticated approaches like Jacobi weighting must be used to ensure convergence~\cite{varnavides-prb}.

\subsection{Collisional Invariants}\label{sec:collisional-invariants}

Since our aim in this \textit{Article} is to investigate hydrodynamic theories of electrons in condensed matter, and hydrodynamic theories are effective theories of conserved quantities in interacting systems, we now turn our attention to conserved quantities within~\cref{eq:bte2}.
Consider a quantity
\begin{align}
    q = \int d^d r\; q_i(\boldsymbol{r}) n_i(\boldsymbol{r}),
\end{align}
where $q_i(\boldsymbol{r})$ is the value of this quantity in state $i$ at position $\boldsymbol{r}$.

Since the left-hand side of~\cref{eq:bte1}, known as the streaming operator, intrinsically conserves carrier quantities, and our collision operator is linearized around a spatially-homogeneous equilibrium distribution, we can investigate the rate of change of $q$ by looking at a translation-invariant system.
In such a system, the time evolution of $q$ may be written with~\cref{eq:bte1} as:
\begin{align}
    \frac{d q}{d t} =  q_i \frac{\partial n_i}{\partial t} = q_i \frac{\partial \Gamma_i}{\partial n_j} \delta n_j,
\end{align}
where we have once again linearized the collision operator.
In order for $q$ to be conserved for all $\delta n_j$, we thus require:
\begin{align}
    q_i \frac{\partial \Gamma_i}{\partial n_j} =0 \quad\,\forall j.
\end{align}
That is, the vector $\{q_i\}$ must live in the left null-space of $\frac{\partial \Gamma_i}{\partial n_j}$.

We can impose this restriction on a non-conservative collision operator by left-projecting out $\{q_i\}$ as
\begin{align}
    R_{q,ik} \equiv \left(\delta_{ij}-\frac{q_i q_j}{\sum_i q_i^2}\right) \frac{\partial \Gamma_j}{\partial n_k} \label{eq:projection},
\end{align}
where time evolution with $R_q$ conserves $q$.
The process generalizes to the case of multiple conserved quantities $q^1,q^2,\dots$, with the only subtlety being that we must first transform these into an orthonormal basis before using~\cref{eq:projection} so that the projection operators commute.

\begin{figure*}
    \centering
    \includegraphics[width=\linewidth]{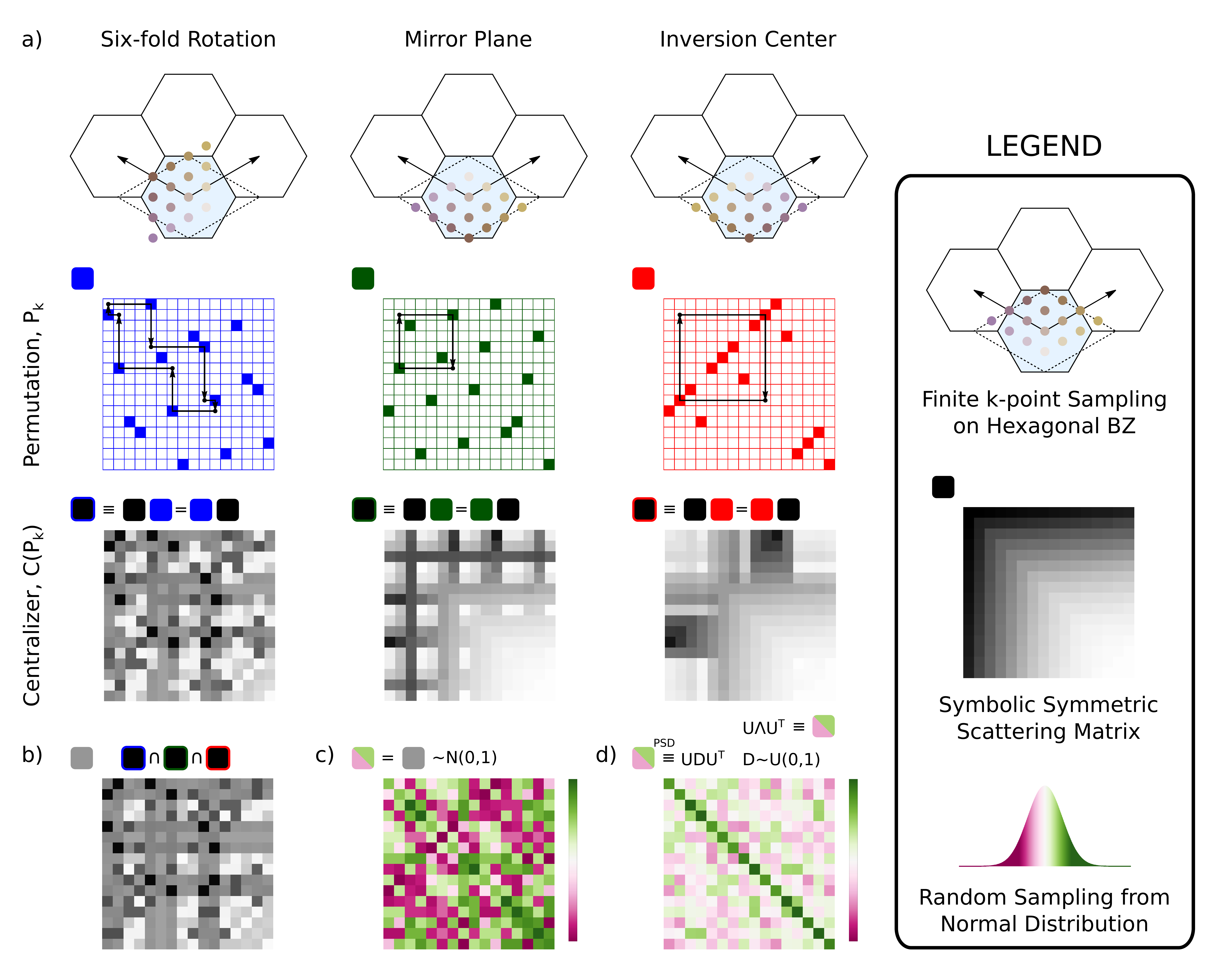}
    \caption{Schematic of procedure to generate physically-plausible collision operators randomly respecting crystal symmetries.
    \textbf{(a)} Finite states in the first Brillouin zone (legend) permuted by each crystal symmetry generator to a new state (top row).
    $N \times N$ permutation matrices corresponding to each symmetry generator (second row). 
    Overlaid arrows show one cycle starting at state $i=2$.
    Starting with a symbolic symmetric scattering matrix (legend), we construct the centralizers for each permutation matrix (the basis elements of the vector space of matrices commuting with each permutation matrix, third row).
    \textbf{(b)} Maximal set of independent components identified by taking the union of the centralizers for each symmetry generator in \textbf{(a)}.
    \textbf{(c)} Independent components assigned by sampling a random-normal distribution with zero mean and unit variance (legend).
    \textbf{(d)} Numeric matrix in \textbf{(c)} is spectrally-decomposed to extract symmetry-preserving eigenvectors, assigned random uniform eigenvalues ensuring positive-definiteness, and re-combined to form the collision operator.
    }
    \label{fig:fig1}
\end{figure*}

Before describing the procedure to generate a physically-plausible state-resolved $R_q$ using arbitrary crystal symmetries, we note that in the case of an isotropic system we can identify the terms in the common dual relaxation-time approximation~\cite{lorentz1905,Callaway1959,deJong1995} with carrier and momentum projection operators:
\begin{align}
    \Gamma[n] \approx &\frac{1}{\tau_{\mathrm{mr}}}\left[-\delta n_i+ \frac{1}{N} \sum_j \delta n_j \right] + \notag \\
    &\frac{1}{\tau_{\mathrm{mc}}}\left[-\delta n_i+ \frac{1}{N} \sum_j \delta n_j\left(1+2\boldsymbol{\hat{v}}_j \cdot\boldsymbol{\hat{v}}_i\right) \right] \label{eq:rta1}.
\end{align}
Here, $\tau_{\mathrm{mr/mc}}$ refers to scalar momentum-relaxing/conserving lifetimes and $N$ is the total number of states.

\Cref{eq:rta1} can be written in the more illustrative form
\begin{align}
        \Gamma[n] \approx \frac{1}{\tau_{\mathrm{mr}}}\left[\mathbb{P}_c -1\right]\delta n + \frac{1}{\tau_{\mathrm{mc}}}\left[\mathbb{P}_m -1\right]\delta n \label{eq:rta2},
\end{align}
where $\mathbb{P}_c$ and $\mathbb{P}_m$ are (isotropic) projectors which ensure carrier number and momentum conservation respectively.

\begin{figure*}
    \centering
    \includegraphics[width=\linewidth]{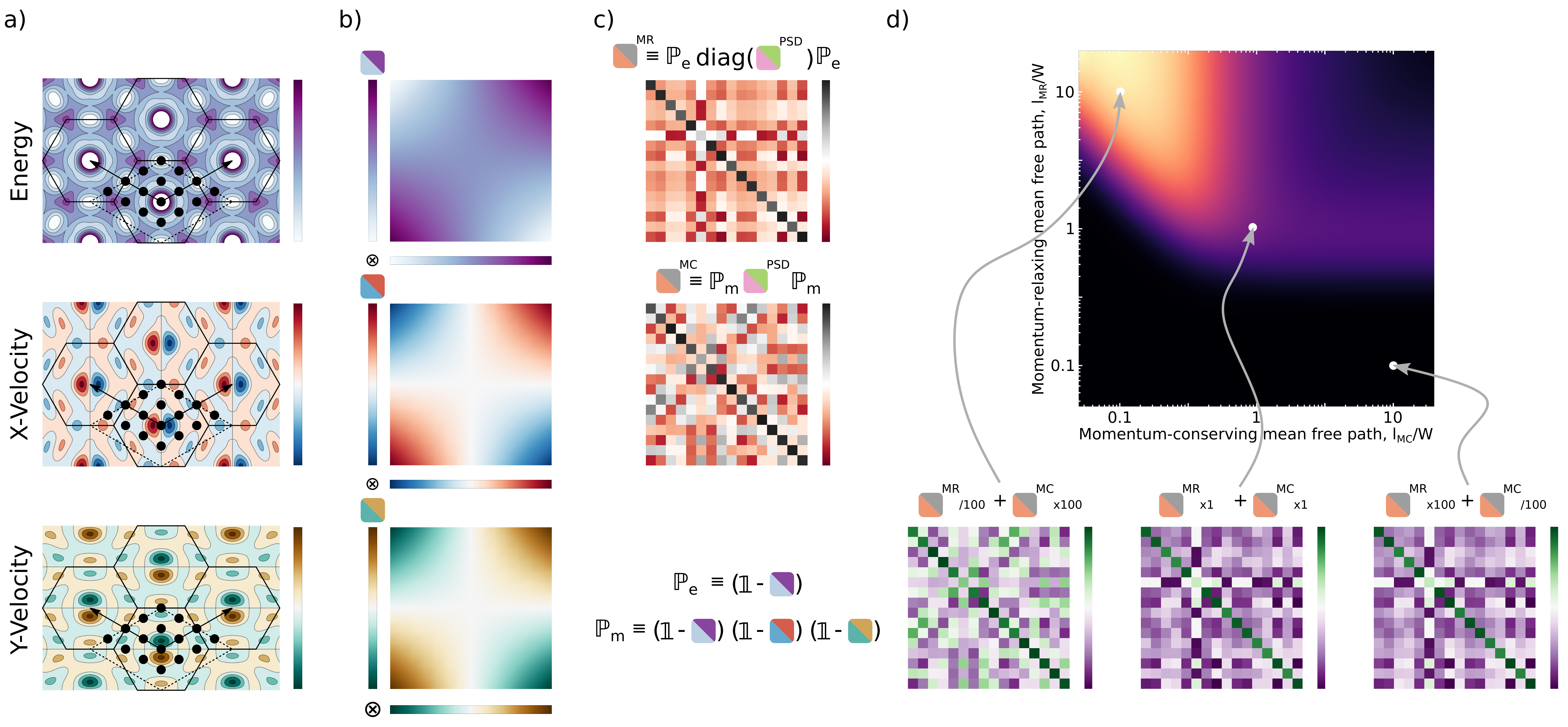}
    \caption{Schematic of procedure to project conserved quantities in physically-plausible collision operators.
    \textbf{(a)} Finite sampling of state-resolved quantities (energy and in-plane group velocities).
    \textbf{(b)} Projection operators constructed for each conserved quantity.
    \textbf{(c)} Momentum-relaxing and momentum-conserving collision operators constructed by projecting out the relevant quantities from the random numeric matrix from~\cref{fig:fig1}d.
    \textbf{(d)} Linear combinations of momentum-conserving and momentum-relaxing collision operators from \textbf{(c)} scaled according to~\cref{eq:norm-mc,eq:norm-mr}.
    }
    \label{fig:fig2}
\end{figure*}

\subsection{Physically Plausible State-Resolved Collision Operators}\label{sec:collision-operators}

Despite the remarkable simplification of the BTE afforded by~\cref{eq:rta1}, and its ubiquity in recent electron hydrodynamic studies~\cite{scaffidi2017,sulpizio_visualizing_2019,vool2021imaging,varnavides2021finitesize}, it offers little insight into how the microscopic scattering details of the collision operator manifest in macroscopic observables.
While \cref{eq:bte3} allows us to go beyond the dual relaxation time approximation, it is often computationally intractable to compute the linearized collision matrix from first principles.
Here, we propose a framework to generate collision operators randomly, constrained by crystal symmetries and conservation laws~\cite{hardy1970}, and scaled according to experimentally-accessible lifetimes:

\begin{enumerate}[label=\roman*)]

    \item We start by specifying the crystallographic point-group of the structure, $\mathcal{G}$, and sampling the first Brillouin zone (BZ) to obtain $N$ discrete states (\cref{fig:fig1}, legend).
    The number of states in the BZ sets our state-space resolution, and is only restricted to be divisible by the lengths of the orbits in the symmetry group of $\mathcal{S}[\mathcal{G}]$.

    \item For each of the structure's $k$ symmetry generators, we construct $N \times N$ permutation matrices $\mathcal{P}_k$ which map one discrete state to another (\cref{fig:fig1}a, middle row).
    Recall that permutation matrices are binary square matrices with exactly one non-zero entry in each row and each column satisfying $\mathcal{P}_k^m = I$, i.e. repeated $m$ applications, where $m$ is the the symmetry order, return the identity matrix (shown graphically by the black permutation paths).
    
    \item Starting with a symmetric symbolic matrix (i.e. with $N\times\left(N+1\right)/2$ independent elements, \cref{fig:fig1} legend),  we construct the centralizer $\mathcal{C}$ for each $\mathcal{P}_k$.
    This is defined as the vector space of all matrices commuting with $\mathcal{P}_k$ (\cref{fig:fig1}a, bottom row):
    \begin{align}
        \mathcal{C} (\mathcal{P}_k) = \left\{ X \in M_{N \times N}(\mathbb{R}) \; \middle| \; \mathcal{P}_k X = X \mathcal{P}_k \right\}.
    \end{align}
    The maximal set of independent components is given by the intersection of these centralizers across permutation matrices (\cref{fig:fig1}b), which is solved as a recursive linear system since $\mathcal{P}_k$ are binary.
    
    \item We sample these independent components from a normal distribution with zero mean and unit variance (\cref{fig:fig1} legend), yielding a random scattering matrix $R_0$ which respects the system's crystal symmetries (\cref{fig:fig1}c).
    
    \item As constructed, $R_0$ may have unphysical (negative) eigenvalues.
    We therefore proceed by enforcing the matrix be positive semi-definite.
    This is done by spectrally decomposing the matrix and replacing its eigenvalues with random-uniform eigenvalues (\cref{fig:fig1}d):
    \begin{align}
        U \Lambda U^T &= R_0 \notag \\
        R_0^{\mathrm{PSD}} &= UDU^T \qquad \mathrm{s.t.}\; D \sim \mathcal{U}(0,1)
    \end{align}

\end{enumerate}

Steps (i-v) above generate a random positive semi-definite collision operator respecting the system's crystal symmetries.

Next, we extend the procedure outlined above to project out conserved quantities of interest.
In particular, we seek to generalize~\cref{eq:rta2} by constructing two state-resolved collision operators $R_{\mathrm{mr}}$ and $R_{\mathrm{mc}}$ where $R_{\mathrm{mr}}$ conserves carriers and $R_{\mathrm{mc}}$ conserves carriers and momentum.
We then construct a family of collision operators with a tunable ratio of momentum-conserving to momentum-relaxing scattering according to:
\begin{align}
    R(\tau_{\mathrm{mc}},\tau_{\mathrm{mr}}) = \tau_{\mathrm{mc}}^{-1} \frac{R_{\mathrm{mc}}}{\left|R_{\mathrm{mc}}\right|} + \tau_{\mathrm{mr}}^{-1} \frac{R_{\mathrm{mr}}}{\left|R_{\mathrm{mr}}\right|} \label{eq:linear-comb}.
\end{align}

\begin{figure*}
    \centering
    \includegraphics[width=0.85\linewidth]{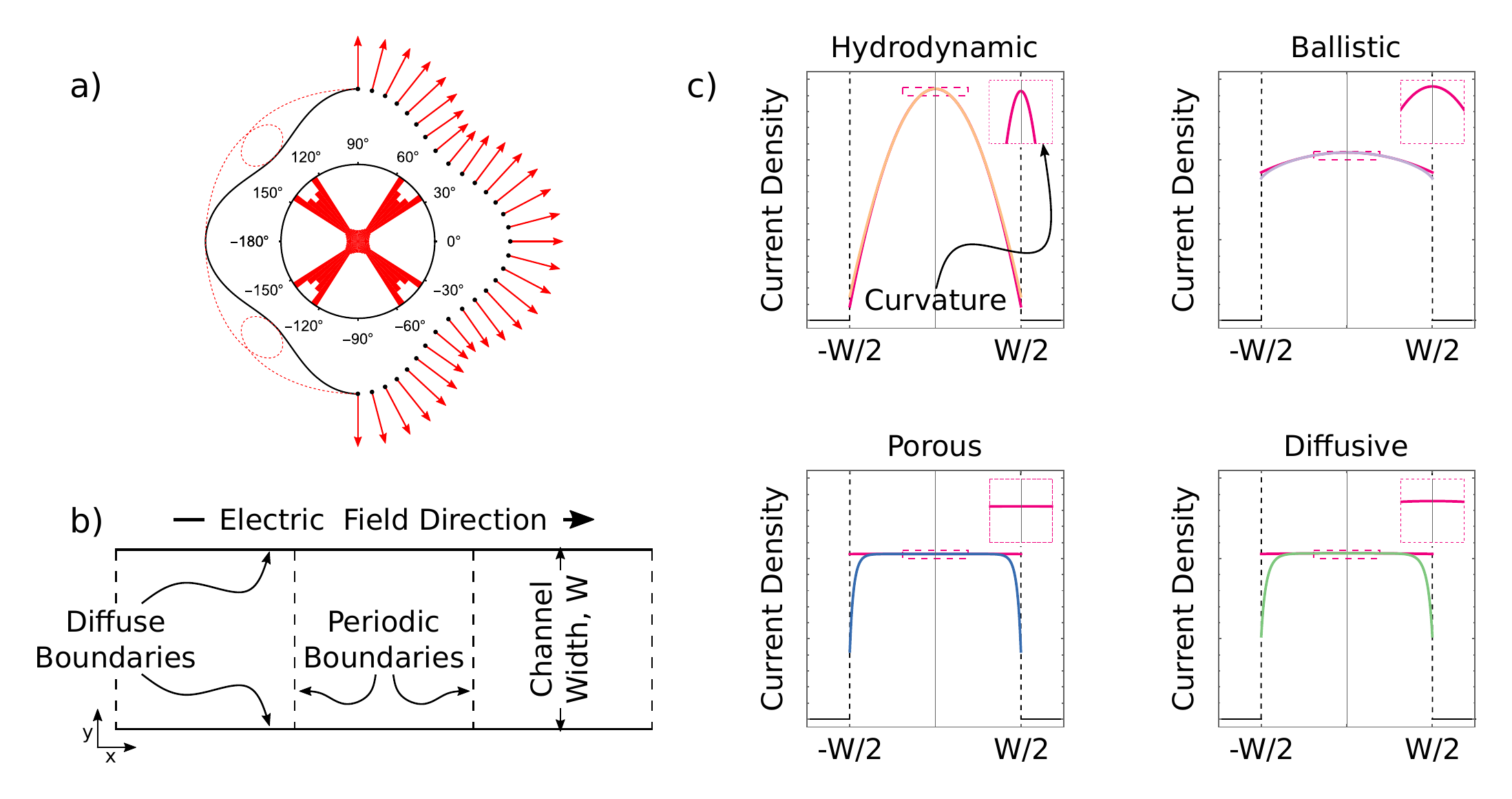}
    \caption{\textbf{(a)} Fermi surface with square $(D_4)$ symmetry (black solid line) parametrized in $N$ discrete steps (black points). 
    Unit magnitude velocities are shown in red arrows.
    Swept over the Fermi wavevector angle, the velocities trace a self-intersecting curve (red dotted line), leading to strongly anisotropic velocity distributions (inset histogram).
    \textbf{(b)} Channel flow geometry of width $W$, with periodic and diffuse boundaries along the $\hat{x}$ and $\hat{y}$ directions respectively.
    \textbf{(c)} Current densities shown for the hydrodynamic (top-left), ballistic (top-right), porous (bottom-left), and diffusive (bottom-right) transport regime limits.
    The curvature obtained by fitting a parabola to the current distribution in the channel center is shown in the inset.}
    \label{fig:fig3}
\end{figure*}

We start with $R_0^{\mathrm{PSD}}$ (\cref{fig:fig1}d) and perform the following steps:
\begin{enumerate}[label=\roman*)]
    \setcounter{enumi}{5}
    \item Define the momentum projection operator as (\cref{fig:fig2}a-c)
    \begin{align}
        \mathbb{P}_m = \prod_k \left(I -u^k \otimes u^k \right),
    \end{align}
    where $u^k$ is the orthonormal basis which spans the same $d+1$-dimensional space as the $\{\chi_i\}$ space spanned by the energy and $d$-components of the group-velocity and $d$ is the spatial dimension (e.g. in 2D $\chi_i = \left\{\epsilon_i, v_i^x, v_i^y\right\}$).
    Note that because we pin carriers to the Fermi surface conserving energy automatically also conserves carriers, but had we not done this we would additionally need to conserve carriers by including the vector $z_i = 1$ in $\chi$ (i.e. each carrier contributes $1$ to the carrier count). 
    
    \item Construct the momentum-conserving collision operator by projecting $R_0^{\mathrm{PSD}}$ on both sides (\cref{fig:fig2}c)
    \begin{align}
        R_{\mathrm{mc}} = \mathbb{P}_m \,R_0^{\mathrm{PSD}}\, \mathbb{P}_m.
    \end{align}
    Projection on the left ensures momentum conservation, and projection on the right ensures $R_{\mathrm{mc}}$ is symmetric.
    The resulting collision operator has $d+1$ zero eigenvalues and spectral radius~$\sim$1.
    
    \item In principle, the maximally momentum-relaxing collision operator is similarly given by defining the operator $\mathbb{Q}_m = I - \mathbb{P}_m$ to project $R_0^{\mathrm{PSD}}$ with on both sides.
    However, the resulting operator only has $d$ non-zero eigenvalues leading to a spectral radius of $N/2-1$.
    While this is physically permissible, it is numerically very challenging~\cite{varnavides-prb}.
    Instead, we approximate the momentum-relaxing collision operator using an anisotropic analogue of the first term in~\cref{eq:rta1} (\cref{fig:fig2}c):
    \begin{align}
        R_{\mathrm{mr}} = \mathbb{P}_{\epsilon}\,\mathrm{diag}\left(R_0^{\mathrm{PSD}}\right) \, \mathbb{P}_{\epsilon},
    \end{align}
    where $\mathbb{P}_{\epsilon}$ is the energy projection operator.
    $R_{\mathrm{mr}}$ now has only one zero eigenvalue and a spectral radius of unity.
    
\end{enumerate}

Steps (i-viii) above generate the two random positive semi-definite operators in~\cref{eq:linear-comb} respecting the system's crystal symmetries and appropriate conservation laws. 
The only remaining steps are to specify appropriate normalizations $|R|$.

\begin{enumerate}[label=\roman*)]
    \setcounter{enumi}{8}
    
    \item We normalize the momentum-conserving collision operator to have an average lifetime of unity by multiplying $R_{\rm mc}$ through by (\cref{fig:fig2}d) 
    \begin{align}
        |R_{\mathrm{mc}}|^{-1} = \frac{1}{N} \left(R_{\mathrm{mc},ii}\right)^{-1} \label{eq:norm-mc}.
    \end{align}
    
    \item By contrast, we normalize the momentum-relaxing collision operator to have a Drude lifetime of unity by multiplying $R_{\rm mr}$ through by (\cref{fig:fig2}d, \cref{app:drude})
    \begin{align}
        |R_{\mathrm{mr}}|^{-1} = \frac{1}{N v_F}\sqrt{\sum_{\alpha \beta}\left(v_j^{\alpha} \left(R_{\mathrm{mr}}\right)^{-1}_{ji}v_i^{\beta} \right)^2} \label{eq:norm-mr}.
    \end{align}
    
\end{enumerate}

Note that for an isotropic system, steps (i-x) fully-specify the collision operator, i.e. there are no independent components, exactly reproducing the dual relaxation-time approximation collision operator~\cref{eq:rta1,eq:rta2}.

\begin{figure*}
    \centering
    \includegraphics[width=0.9\linewidth]{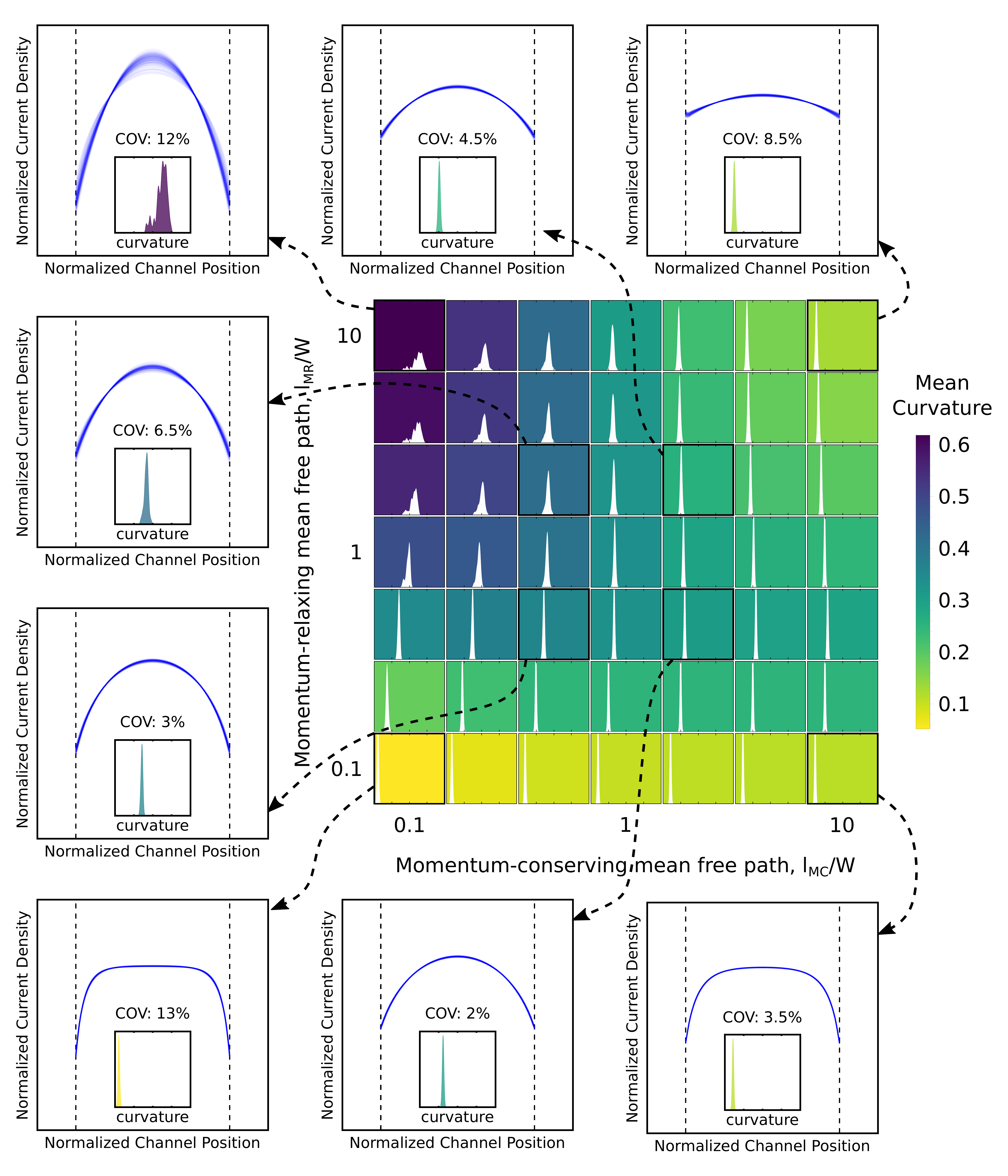}
    \caption{
    Current-density variability results for a system with square ($D_4$) symmetry.
    Distribution of curvatures for $K=100$ random collision operator instances on a grid of normalized momentum-conserving and momentum-relaxing mean free paths, using a common scale highlighting the increasing curvature variability in the hydrodynamic regime.
    Each grid tile's background is color-coded according to its mean curvature.
    Callouts plot the $K=100$ normalized current densities for each random instance, for the four transport regime limits (top-right: ballistic, top-left: hydrodynamic, bottom-left: porous, bottom-right: diffusive), as well as four more experimentally-accessible transport regimes symmetric about $l_{\mathrm{mr}}=l_{\mathrm{mc}}=W$.
    Insets show zoomed-in curvature distributions, together with the distribution's coefficient of variability $\mathrm{COV} (\%)=\sigma/\mu \times 100$.
    }
    \label{fig:fig4}
\end{figure*}

\section{Results}\label{sec:results}

\subsection{Current Density Variability}\label{sec:current-density}

We now apply the formalism developed in~\cref{sec:form} to investigate the sensitivity of hydrodynamic electron channel flow to the details of the collision operator. 
We restrict ourselves to scattering between states on the Fermi surface (\cref{fig:fig3}a) and work in a two-dimensional channel, with periodic and diffuse boundary conditions~\cite{deJong1995,sulpizio_visualizing_2019,vool2021imaging} along the $\hat{x}$ and $\hat{y}$ directions respectively (\cref{fig:fig3}b).
A weak electric field $E \hat{x}$ is applied to drive the current. 
We choose a unit-magnitude velocity normal to the Fermi surface, and fix the units such that a lifetime of $\tau=1$ corresponds to a mean-free path equal to the channel width.
This allows us to parameterize our collision operators in the non-dimensionalized space of $\left(l_{\mathrm{mc}}/W,l_{\mathrm{mr}}/W\right)$, where $l_{\mathrm{mc/mr}}$ are the momentum-conserving/relaxing mean free paths, respectively, and $W$ is the width of the channel.

For each point in a grid of $\left(l_{\mathrm{mc}}/W,l_{\mathrm{mr}}/W \right)$, we generated $K=100$ random collision operators.
For each collision operator we solve~\cref{eq:bte3} to obtain the spatially resolved carrier population $\delta n_i(\boldsymbol{r})$, from which we derive the current density profile.
We then fit a parabola to the current density profile to obtain the current density curvature (\cref{fig:fig3}c).

We begin with a system with square ($D_4$) symmetry (\cref{fig:fig4}).
Each square in the grid corresponds to one choice of $\left(l_{\mathrm{mc}}/W,l_{\mathrm{mr}}/W \right)$ and colored by the mean curvature of all $K=100$ collision operators we generated with those parameters.

We see increasing curvature towards long momentum-relaxing and short momentum-conserving mean-free paths, signalling the onset of hydrodynamics.
This is consistent with the transport-regime diagrams characteristic of the dual relaxation-time approximation~\cite{sulpizio_visualizing_2019,vool2021imaging,varnavides2021finitesize}.

The distribution of curvatures is overlaid in white for each square, highlighting that the variability between samples is larger in the hydrodynamic regime (upper-left) than in the diffusive regime (lower-right) or ballistic (upper-right).
This may also be seen in the current density plots surrounding the grid.

We use the coefficient of variability $\mathrm{COV} (\%)= \sigma/\mu \times 100$, where $\mu$ and $\sigma$ are the distribution's mean and standard deviation respectively, as a normalized measure of variance and find that, even in the hydrodynamic regime, the macroscopic current density observable only exhibits a $\leq $12\% variability between samples.
If we instead restrict ourselves to the more experimentally-accessible regime of $l_{\mathrm{mr}}\sim l_{\mathrm{mc}}\sim W$ we find that the variability is on the order of $\sim$5\%, which is smaller than typical experimental uncertainties.

We find that the current density profile is most sensitive to the details of the scattering processes in the hydrodynamic limit, though even there this sensitivity is minimal, and the current density profile is primarily a measure of the relative rates of momentum-conserving and momentum-relaxing scattering processes.

\begin{figure*}
    \centering
    \includegraphics[width=\linewidth]{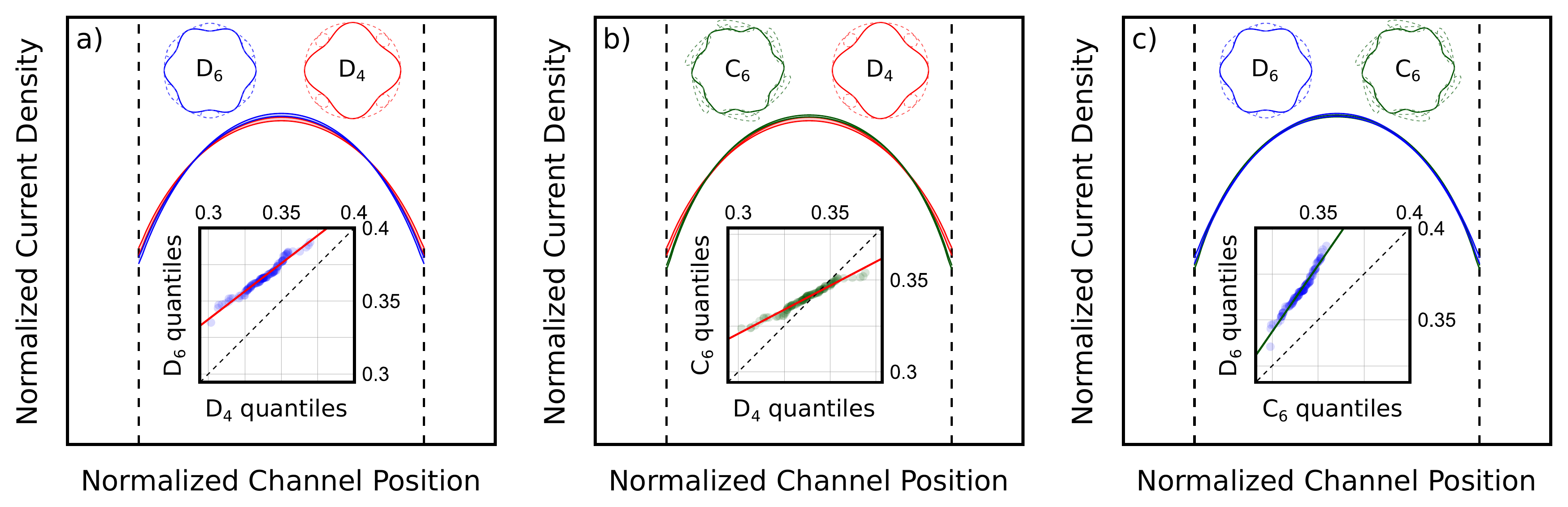}
    \caption{Comparison of normalized current-density profiles for different crystallographic point groups for $l_{\mathrm{mr}}=l_{\mathrm{mc}}=W$.
    \textbf{(a)} 95\% confidence intervals for $K=100$ random instances for systems with square ($D_4$, red) and hexagonal ($D_6$, blue) crystal symmetries.
    Inset shows a QQ-plot of the hexagonal calculation's curvature quantiles against those of the square calculation.
    The reasonable fit suggests the distributions have similar variance, with the rigid offset indicating the hexagonal calculation has larger curvatures.
    Similar comparison as in \textbf{(a)}, comparing \textbf{(b)} a six-fold ($C_6$, green) system with a square ($D_4$, red) system and \textbf{(c)} a hexagonal ($D_6$, blue) system with a six-fold ($C_6$, green) system.
    }
    \label{fig:fig5}
\end{figure*}

Finally, we illustrate the generality of our framework by investigating two additional crystal systems, those with hexagonal ($D_6$) and six-fold ($C_6$) symmetry.
\Cref{fig:fig5} summarizes the results by plotting pairwise-comparisons between different symmetries in the different panels for one point in the $(l_{\mathrm{mc}}/W,l_{\mathrm{mr}}/W)$ grid, the experimentally-accessible regime $l_{\mathrm{mr}} = l_{\mathrm{mc}} = W$.
Each panel shows the current density profiles.
To understand the differences between these distributions we turn to the quantile-quantile plots (insets).
Each quantile-quantile plot matches each percentile curvature of one symmetry with the corresponding percentile of the other.
Importantly, in the inset of (\cref{fig:fig5}a) we see that the curvatures of $D_6$ are systematically higher than those of $D_4$, which results from the anisotropy of the Fermi velocity~\cite{varnavides2021finitesize}.

\subsection{Viscosity Tensor Variability}\label{sec:viscosity-tensor}

A common approach to avoid the computational cost of solving~\cref{eq:bte3} is to instead take the first moment of the BTE to recover the electronic Stokes equation~\cite{Levitov2016,varnavides2020electron,scaffidi2017}:
\begin{align}
    \partial_t u_{\alpha} = \partial_{\beta} T^{(v)}_{\beta \alpha} -\partial_{\alpha} P - R_{\alpha \beta} u_{\beta} \label{eq:ns}
\end{align}
where $u$ is the electron fluid velocity (current density), $P$ is the pressure (related to electrochemical potential $\Phi$ through $P = e\int_{n_0}^n \Phi(n')dn' \approx e(n-n_0) \Phi$), $R_{\alpha \beta}$ is a positive semi-definite rank-2 tensor proportional to the momentum-relaxing rate, $T^{(v)}_{\alpha \beta}$ is the viscous contribution to the stress (momentum flux) tensor, and Greek letter subscripts correspond to cartesian directions.
The stress is related to the fluid velocity to linear order by the rank-4 electronic viscosity tensor, $A_{\alpha \beta \gamma \zeta}$:
\begin{align}
    T^{(v)}_{\alpha \beta} = A_{\alpha \beta \gamma \zeta} \partial_{\zeta} u_{\gamma}. \label{eq:viscosity}
\end{align}

As with all physical properties in crystals, the viscosity tensor is constrained by i) ``intrinsic'' tensor symmetries, ii) crystal symmetries, and iii) thermodynamic stability.
Due to the rich landscape of viscous effects such as Hall-viscosity and vorticity coupling enabled by preferred directions inside crystals~~\cite{Avron1998,Bradlyn2020,Epstein2020,varnavides2020electron}, the solution of~\cref{eq:ns} hinges on accurate calculation of the viscosity tensor.

In Appendix~\ref{app:visc} we derive a method to calculate the viscosity tensor given a collision operator.
In this section we evaluate the viscosity and its variability using the same random collision operator instances from~\cref{sec:current-density}.
\Cref{table:1} summarizes the normalized viscosity means and standard deviations using the same random collision operator instances as in~\cref{sec:current-density}.
Note that our choice to produce symmetric collision operators in~\cref{sec:collision-operators} implies microscopic reversibility and thus precludes any Hall-viscosity components.
Similarly, the incompressibility condition we used in~\cref{app:visc} means that we cannot measure the components of the viscosity that couple to compression, i.e. we obtain $A_{ij11}=-A_{ij22}$.
We see that the viscosity tensor varies substantially between collision operators of fixed $\tau_{\rm MR/MC}$, with coefficients of variability $> 50\%$.
This is in contrast to the minimal variation we saw in the current density profiles in~\cref{sec:current-density}, and indicates that the viscosity tensor is sensitive to the microscopic details of the collision operator.

\section{Conclusions}\label{sec:conclusions}

We present a theoretical and computational formalism to generate physically plausible collision operators respecting a system's crystal symmetries and conservation laws.
We use this formalism to quantify the sensitivity of macroscopic observables to microscopic interaction details in electron hydrodynamics.
We find that the detailed transition rates between electronic states produce small corrections, of order $\sim$10\%, to current density profiles in 2D channel flow.
Rather, the curvature in the middle of the channel is set primarily by the size of the geometry, and the ratio of momentum-relaxing to momentum-conserving lifetimes.
This has several key implications.

First, for simple geometries such as channel flow, it should be possible to accurately predict the current density profile armed only with the knowledge of the momentum-relaxing and momentum-conserving lifetimes.
As lifetime data are generally more experimentally accessible and readily predicted from first principles than collision operators, this represents a significant simplification to such calculations while retaining the full spatial and state resolution of the BTE.
Conversely, this suggests that microscopic interaction details are difficult to probe with current density measurements since~\cite{lukin2015,lukin2019}, at fixed lifetimes, the variance in current density profiles across random collision operator instances is small.
More spatially-complex electronic flows, such as rotational flow in a Corbino disk geometry~\cite{varnavides2020electron} and vortices in a double chamber geometry~\cite{aharonsteinberg2022direct}, may fill this gap.
Secondly, we find that the viscosity tensor -- a key input parameter in modeling these electron fluids using the electronic Stokes equation~\cite{Levitov2016,scaffidi2017,varnavides2020electron} -- is significantly more sensitive to the details of the collision operator, with variance across samples larger than 50\%.
Thus the action of viscosity on the channel flow resistance provides an experimental probe of the underlying scattering processes.

Taken together, our observations suggest several paths towards probing momentum-conserving interactions in electron fluids.
First, spatially resolved current density measurements are a powerful tool to assess the regime of transport (namely, hydrodynamic or not) and the relative rates of momentum-conserving and momentum-relaxing scattering processes, but are not as informative on the details of those scattering processes.
Secondly, viscosity measurements such as channel flow resistance and/or Corbino disk voltage drops~\cite{varnavides2020electron}, which are only meaningful in the hydrodynamic limit, can probe the details of the collision operator.
Thus the two work in conjunction: current density measurements allow one to identify a hydrodynamic system, and viscosity measurements allow one to probe the underlying scattering physics.

Finally, we note that our formalism for producing random collision operators can be readily adapted to other conserved quantities (for example spin), symmetries, and random distributions.
Thus it provides a fast and adaptable tool for modelling scattering in a wide range of contexts.
\section*{Acknowledgments}\label{sec:acknowledments}

This work was supported by the Quantum Science Center (QSC), a National Quantum Information Science Research Center of the U.S. Department of Energy (DOE).
This research used resources of the Oak Ridge Leadership Computing Facility, which is a DOE Office of Science User Facility supported under Contract DE-AC05-00OR22725 as well as the resources of the National Energy Research Scientific Computing Center, a DOE Office of Science User Facility supported by the Office of Science of the U.S. Department of Energy under Contract No. DE-AC02-05CH11231.
The Flatiron Institute is supported by the Simons Foundation.
P.N. is a Moore Inventor Fellow and gratefully acknowledges support through Grant No. GBMF8048 from the Gordon and Betty Moore Foundation. 

\appendix
\section{Drude Model Lifetime} \label{app:drude}

Here we relate lifetimes in our formalism to those given by the Drude model of conductivity.
The linearized steady state BTE with an electric field is given by
\begin{align}
e \boldsymbol{v}_i\cdot\boldsymbol{E}\left(-\frac{dn_0}{d\epsilon}\right)_i=M_{ij}n_j,
\end{align}
where $n_0$ is the Fermi-Dirac distribution, $\epsilon$ is energy, and $\boldsymbol{E}$ is the gradient of the electrochemical potential.
We can solve for $n_j$ to find
\begin{align}
n_j = M_{ji}^{-1} \left(e \boldsymbol{v}_i\cdot\boldsymbol{E}\left(-\frac{dn_0}{d\epsilon}\right)_i\right)
\label{eq:nj}
\end{align}
where $M_{ji}^{-1}$ is the pseudoinverse of $M_{ij}$.

The momentum and current density of carriers is given by
\begin{align}
    \boldsymbol{p} &= m_e n_i \boldsymbol{v}_i \\
    \boldsymbol{j} &= e n_i \boldsymbol{v}_i = \frac{e}{m_e}\boldsymbol{p}
\end{align}
Which we can relate to the electric field using the steady state distribution (equation~\ref{eq:nj}):
\begin{align}
\boldsymbol{j} = e^2 \boldsymbol{v}_j M_{ji}^{-1} \boldsymbol{v}_i \cdot\boldsymbol{E}\left(-\frac{dn_0}{d\epsilon}\right)_i.
\end{align}
The conductivity tensor is then given by
\begin{align}
\sigma_{\alpha \beta} = e^2 v_j^\alpha M_{ji}^{-1} v_i^\beta \left(-\frac{dn_0}{d\epsilon}\right)_i.
\label{eq:sigma}
\end{align}

At low temperatures $\frac{dn_0}{d\epsilon}$ restricts contributions to $\sigma$ to come from scattering near the Fermi surface.
Restricting sums over $i$ and $j$ to run over states near the Fermi surface then we can replace $\frac{dn_0}{d\epsilon}$ with the density of states, $g$ :
\begin{align}
\sigma_{\alpha \beta} = -\frac{e^2 g}{N}\left(v_j^\alpha M_{ji}^{-1} v_i^\beta\right),
\end{align}
where $N$ is the number of states we sample on the Fermi surface.
On a spherical Fermi surface with a quadratic dispersion relation this simplifies to:
\begin{align}
\sigma_{\alpha \beta} = -\frac{n e^2}{m_e v_F^2 N}\left(v_j^\alpha M_{ji}^{-1} v_i^\beta\right),
\end{align}
where $v_F$ is the Fermi velocity.

In the relaxation time approximation, $M_{ji}^{-1} = -\tau \delta_{ij}$, so
\begin{align}
\sigma_{\alpha \beta} = \frac{n e^2 \tau }{m_e}\sum_{i,\alpha} \frac{\left(v_i^\alpha\right)^2}{N v_F^2} \delta_{\alpha \beta} = \frac{n e^2 \tau }{m_e}\delta_{\alpha \beta},
\end{align}
which is we identify as the Drude conductivity.
For systems outside relaxation time approximation then, we can instead ascribe a mean lifetime
\begin{align}
\tau_{\rm MR}^2 \equiv \frac{1}{N v_F^2} \sum_{\alpha \beta} \left(v_j^\alpha M_{ji}^{-1} v_i^\beta\right)^2,
\end{align}
where we've taken a Frobenius norm of the conductivity tensor to arrive at a scalar lifetime.
\begin{table*}
\centering
\begin{tabular}{c c c}
 \textbf{Hexagonal Viscosity Tensor,} $A_{\alpha \beta \gamma \zeta}^{D_6}$ & & \textbf{Square Viscosity Tensor,} $A_{\alpha \beta \gamma \zeta}^{D_4}$ \\ 
 \hline
 & \\
$\left(
\begin{array}{cc}
 \left(
\begin{array}{cc}
 0.35\pm 0.18 & 0. \\
 0. & -0.35\pm 0.18 \\
\end{array}
\right) & \left(
\begin{array}{cc}
 0. & 0.35\pm 0.18 \\
 0.35\pm 0.18 & 0. \\
\end{array}
\right) \\
 \left(
\begin{array}{cc}
 0. & 0.35\pm 0.18 \\
 0.35\pm 0.18 & 0. \\
\end{array}
\right) & \left(
\begin{array}{cc}
 -0.35\pm 0.18 & 0. \\
 0. & 0.35\pm 0.18 \\
\end{array}
\right) \\
\end{array}
\right)$ & &
$\left(
\begin{array}{cc}
 \left(
\begin{array}{cc}
 0.17\pm 0.14 & 0. \\
 0. & -0.17\pm 0.14 \\
\end{array}
\right) & \left(
\begin{array}{cc}
 0. & 0.5\pm 0.4 \\
 0.5\pm 0.4 & 0. \\
\end{array}
\right) \\
 \left(
\begin{array}{cc}
 0. & 0.5\pm 0.4 \\
 0.5\pm 0.4 & 0. \\
\end{array}
\right) & \left(
\begin{array}{cc}
 -0.17\pm 0.14 & 0. \\
 0. & 0.17\pm 0.14 \\
\end{array}
\right) \\
\end{array}
\right)$
\end{tabular}
\caption{2D Viscosity tensors computed using $K=100$ random collision operator instances for hexagonal ($D_6$) and square ($D_4$) systems.
Entries show the normalized mean $\pm$ standard deviation, using the normalization $|A|^2 = A_{\alpha \beta \gamma \zeta} A_{\alpha \beta \gamma \zeta}=1$.
}
\label{table:1}
\end{table*}

\section{Extracting Viscosities} \label{app:visc}

In this section, we derive the viscosity tensor within the BTE framework introduced in~\cref{sec:bte} using a simple carrier distribution perturbation~\cite{Simoncelli2019,PhysRev.109.1486}.

We work in the limit of perfect momentum conservation, and begin with a perturbed carrier distribution of the form:
\begin{align}
    \delta n_{i,1}(\boldsymbol{r}) = \boldsymbol{v}_i \cdot \boldsymbol{u}(\boldsymbol{r}), \label{eq:pert-1}
\end{align}
where $u$ is a background drift velocity field with shears $\partial_{\zeta} u_{\gamma}$.
Because collisions conserve momentum, $\delta n_{i,1}$ is a null eigenvector of the collision operator.
As such, at steady-state, we must add an additional perturbation on top of~\cref{eq:pert-1} to satisfy~\cref{eq:bte2}:
\begin{align}
    -\boldsymbol{v}_i \cdot \nabla \left( \delta n_{i,1} + \delta n_{i,2} \right) =  \frac{\partial \Gamma_i}{\partial n_j}\delta n_{j,2}.
\end{align}

Next, we assume $\boldsymbol{u}(\boldsymbol{r})$ has a constant gradient, and thus $-\boldsymbol{v}_i \cdot \nabla \delta n_{i,1}$ is spatially homogeneous.
This means $\delta n_{i,2}$ is also spatially homogeneous, so we can solve for the perturbation $\delta n_{j,2}$ using the linear system:
\begin{align}
    -\boldsymbol{v}_i \cdot \nabla \delta n_{i,1} = \frac{\partial \Gamma_i}{\partial n_j}\delta n_{j,2}.\label{eq:lin}
\end{align}
The stress tensor is then given by:
\begin{align}
    T^{(v)}_{\alpha \beta} = v_{i,\alpha} v_{i,\beta}\; \delta n_{j,2} \label{eq:viscous-flux}.
\end{align}
This allows us to reconstruct the viscosity tensor $A_{\alpha \beta \gamma \zeta}$ by sampling~\cref{eq:viscous-flux} for various shears $\delta_{\zeta} u_{\gamma}$.

A subtlety here is that in steady state we need to enforce carrier conservation, which for many systems amounts to requiring the flow be incompressible.
To see this we sum the steady state linearized BTE~\cref{eq:bte2} over states and find
\begin{align}
    \sum_i     \left(\frac{\partial \Gamma_i}{\partial n_j} + \delta_{ij} \boldsymbol{v}_i \cdot \nabla_{\boldsymbol{r}} \right)\delta n_j(\boldsymbol{r}) = \sum_i S_i,
\end{align}
If the source term $S$ conserves carriers then the right-hand side vanishes.
Following the reasoning in Section~\ref{sec:collisional-invariants}, we construct collision operators which also conserve carriers to find
\begin{align}
    \sum_i \boldsymbol{v}_i \cdot \nabla_{\boldsymbol{r}} \delta n_i(\boldsymbol{r}) = 0.
\end{align}
Since $\delta n_{i,2}$ is spatially uniform, this becomes
\begin{align}
\sum_i  \boldsymbol{v}_i \cdot \nabla_{\boldsymbol{r}}\delta n_{i,1}(\boldsymbol{r}) = 0.
\end{align}
Inserting~\cref{eq:pert-1} and assuming the group velocities are spatially uniform we find
\begin{align}
\sum_i {v}_i^{\alpha} v_i^\beta \partial_\alpha u_\beta = 0.
\end{align}
For the systems we study here, with $D_4$, $D_6$, and $C_6$ symmetry this amounts to requiring that the flow be incompressible ($\nabla\cdot\boldsymbol{u}=0$), but for more general symmetries the constraint may be more complex.
This incompressibility constraint fundamentally arises from carrier conservation and linearizing the collision operator: if we allowed nonlinear contributions to the collision operator we could obtain steady state flows with compression.

Our approach then is to construct $d^2 - 1$ orthogonal shear tensors with vanishing divergence.
For example, in 2D we use:
\begin{align}
    \delta_{\zeta} u_{\gamma}= \left\{ 
    \begin{pmatrix} 0 & 1 \\ 0 & 0\end{pmatrix}, \;
    \begin{pmatrix} 0 & 0 \\ 1 & 0\end{pmatrix}, \;
    \begin{pmatrix} 1 & 0 \\ 0 & -1\end{pmatrix}
    \right\}.
\end{align}
For each of these we solve~\cref{eq:lin} to calculate the stress tensor~\cref{eq:viscous-flux}, and reconstruct the viscosity tensor as the linear map between these.

\bibliography{refs}

\end{document}